# Probing the adsorption/desorption of amphiphilic polymers at the air-water interface during large interfacial deformations

C. Tregouet[1,2], T. Salez[3,4], N. Pantoustier[1], P. Perrin[1], M. Reyssat[2] and C. Monteux[1,4]

1. Laboratoire Sciences et Ingénierie de la Matière Molle, PSL Research University, UPMC Univ Paris 06, ESPCI Paris, UMR 7615 CNRS, 10 rue Vauquelin, 75231 Paris cedex 05, France

2. UMR CNRS Gulliver 7083, ESPCI Paris, PSL Research University, 75005 Paris, France

3. Univ. Bordeaux, CNRS, LOMA, UMR 5798, F-33405 Talence, France

4. Global Station for Soft Matter, Global Institution for Collaborative Research and Education, Hokkaido University, Sapporo, Hokkaido 060-0808, Japan

cecile.monteux@espci.fr

**Abstract**

Hydrophobically modified polymers are good candidates for the stabilization of liquid interfaces thanks to the high anchoring energy of the hydrophobic parts. In this article we probe the interfacial anchoring of a series of home-made hydrophobically modified polymers of controlled degree of grafting by studying their behavior upon large area dilations and compressions. By comparing the measured interfacial tension to the one that we expect in the case of a constant number of adsorbed monomers, we are able to deduce whether desorption or adsorption occurs during area variations. We find that the polymer chains with the longest hydrophobic grafts desorb for larger compressions than the polymers with the shortest grafts, because of their larger desorption energy. Furthermore, we observe more desorption for polymers with the highest grafting densities. We attribute this counter intuitive result to the fact that for high grafting densities, the length of the polymer loops is shorter, hence the elastic penalty upon compression is larger for these layers, leading to a faster desorption. The dilatation experiments reveal that the number of adsorbed anchors remains constant in the case of chains with low grafting density while chains with the highest degree of grafting seem to show some degree of adsorption during the dilatation. Therefore for these highly grafted chains there may be unadsorbed grafts remaining in the vicinity of the interface, which may adsorb quickly to the interface upon dilatation.

**Introduction**

Polymers are widely used as interfacial stabilizers, emulsifiers or suspension dispersants. As they adsorb at liquid/liquid or solid/liquid interfaces, some monomers adsorb into trains while some monomers form loops or tails extending into the liquid, which provide a steric protection for droplets or particles against coalescence or flocculation. Although the desorption energy of one monomer is of the order of $k_BT$, the overall desorption energy of one polymer chain with $N>>1$ adsorbed monomers is of the order of $Nk_BT$, hence the adsorption of a polymer chain is irreversible. However, the low anchoring energy per monomer enables quick exchanges of monomers between adsorbed trains and loops or tails. Amphiphilic polymers, composed of hydrophilic and hydrophobic parts, such as random and block copolymers or polymers grafted with hydrophobic anchors adsorb more strongly to interfaces with aqueous solutions than homopolymers thanks to the high anchoring energy of the hydrophobic anchors. The structure of amphiphilic polymer layers adsorbed at solid and liquid interfaces has been the object of theoretical and experimental research in the 90s [1]. For block copolymers, it was predicted that the hydrophobic parts form 2D coils while the hydrophilic parts form 3D coils swollen in the bulk solution [2, 3]. At high surface polymer concentration, the hydrophilic parts stretch perpendicularly to the interface thus forming a brush [2–9]. Brushes are also expected for other amphiphilic polymers such as telechelic [10], end-functionalized [11] or hydrophobically modified polymers [12–17], for which hydrophobic grafts adsorb to the interface and hydrophilic parts stretch into the bulk solution.

The dynamical properties of amphiphilic polymer layers under an interfacial compression [18-22] has attracted much less attention than their structure, even though it is relevant in foaming or emulsification processes as well as for foam and emulsion stability [23,24]. For example, during bubble rearrangements in a foam, which sometimes leads to bubble coalescence, bubble interfaces and thin liquid films can be strongly stretched or compressed. A fine understanding of how the polymer architecture influences the dynamical properties of the interfacial layers would help rationalizing the design of polymer molecules for foams and emulsions with improved stability. The pendant drop method [7, 20-22] or Langmuir trough experiments [6,10] are useful methods to investigate the behavior of adsorbed layers during a reduction or increase of the surface area. In the case of amphiphilic diblock or triblock copolymers such as poly(ethylene oxide) – poly(styrene), PEO-PS, or poly(propylene oxide) – poly(ethylene oxide) - poly propylene oxide) copolymers, PPO-PEO-PPO or telechelic polymers, a mushroom to brush transition was observed upon large compression of the layers [4, 6, 7, 10], as initially adsorbed PEO monomers, desorb from the interface to form stretched loops while the more hydrophobic blocks, PPO or PS remain adsorbed at the interface.

Hydrophobically modified polymer molecules, such as PAAH-$\alpha$-$C_n$ composed of a poly(acrylic acid), PAAH, backbone, covalently and randomly grafted with hydrophobic alkyl anchors, $C_n$, with n the number of carbon atoms on the alkyl chain and $\alpha$ the mole fraction of grafted monomers, are known to be very efficient foam and emulsion stabilizers [19, 25–27]. Moreover their chemical architecture —molar mass, number of grafts, degree of ionization and length of grafted alkyl chain— can be easily modified, therefore they are good candidates to investigate how chemical parameters influence the interfacial compression and dilatation of amphiphilic polymers. In a previous article, we have investigated the adsorption dynamics of a series of PAAH-$\alpha$-$C_n$, [17]. We found that at short times, the surface tension is lower for an increasing degree of grafting, while at longer times, as the adsorption dynamics becomes logarithmic, the evolution of the surface tension becomes slower as the degree of grafting increases. Our results were consistent with the formation of a polymer brush at the interface, with PAAH loops extending in solution between adsorbed $C_n$ hydrophobic anchors. We suggested a coarse-grained model where the adsorption dynamics at long times is limited by a free energy-barrier, which results from both the deformation of the incoming polymer coils and the deformation of the adsorbed brush. As the grafting degree increases, the number of monomers between $C_n$ anchors decreases, and hence the length of the loops decreases. This leads to a thinner brush and less stretching of the adsorbing chains and thus to a faster adsorption.

In the present article, we study behavior of adsorbed layers of PAAH-$\alpha$-$C_n$ upon large interfacial compressions and dilatations. By comparing the measured interfacial tension to the one that we expect in the case of a constant number of adsorbed monomers, we deduce that the polymer chains with the longest hydrophobic grafts desorb for larger compressions than the polymers with the shortest grafts, because of the larger desorption energy. Furthermore, counter intuitively we observe more desorption upon compression for higher grafting densities. We attribute this counter intuitive result to the fact that for high grafting densities, the length of the loops is shorter. Hence upon compression, the elastic penalty is larger for these layers, leading to a faster desorption. The dilatation experiments reveal that the number of adsorbed anchors remains constant in the case of chains with low grafting density while chains with the highest degree of grafting seem to show some degree of adsorption during the dilatation. Therefore for these highly grafted chains there may be unadsorbed grafts remaining in the vicinity of the interface, which may adsorb quickly to the interface upon dilatation.

**Materials** We use a series of hydrophobically modified poly(acrylic acid), consisting in a PAA backbone (poly(acrylic acid)) of molar mass of $M_W$ = 100 000 g/mol, covalently grafted with a grafting density $\alpha$ of hydrophobic alkyl grafts of length n. The PAA backbone is provided by Polysciences Inc and the grafting reaction was done according to Ref. [28]. In our work, $\alpha$ ranges between 0.5/100 and 5/100 and $n$ = 8 or 12.

To prepare the polymer solutions, the polymer chains are dissolved at a concentration of 1% w/w for 60 hours. The pH is adjusted to pH=3 with a few milliliters of molar hydrochloric acid solution to obtain conditions in which the PAAH chains do not bear any negative charges. The solutions are then filtered through 0.45 µm PVDF, poly(vinylidene fluoride) membranes. The solutions are stored at a temperature between 4 and 6 °C. The solutions are kept at ambient temperature for 12 hours prior to each experiment. The solutions are used within the seven days after the polymer dissolution.

**Measurements**. To probe the adsorption dynamics of the polymer layers, we use the rising bubble method (Tracker from Teclis, France) which enables the measurement of the surface tension as a function of time by fitting the profile of a bubble to the Laplace equation. Details concerning the principle of this technique can be found in Ref. [29]. Briefly, a fresh millimetric air bubble is formed in the polymer solution and the polymer molecules are left to adsorb at the surface for 3600 seconds. The temperature is maintained constant at 20°C both in the vessel and in the syringe. Then the bubble surface is either reduced or increased by either sucking out air from the bubble or injecting air into the bubble. The interfacial tension and area of the bubble are recorded as a function of time during the compression or dilation of the bubble. The compression or the dilation rates, defined as the relative area variations per unit time, are either 0.007 s$^{-1}$ or 0.1 s$^{-1}$.

# Results

## *Large compressions of the adsorbed layers*

To probe the behavior of the polymer layers under compression, we deflate the air bubble and we record the area and surface tension for PAAH-α–$C_n$ with varying grafting degrees and graft lengths. Figure 1 shows the variation of the surface tension as a function of the deformation $\epsilon$, defined as the relative change in surface area, $\epsilon = \frac{A_i - A}{A_i}$, where $A_i$ is the initial area, and $A$ the area at the considered time. We observe that the surface tension decreases during a compression which means that the surface density of adsorbed polymer chains increases.

If the number of chains remains constant, *ie* the chains do not desorb upon compression, we can easily predict the evolution of the polymer surface excess, $\Gamma$, as a function of the initial surface excess, $\Gamma_i$, and the deformation $\epsilon$:

$$\Gamma = \frac{N_i^{chains}}{A} = \frac{N_i^{chains}}{A_i} \cdot \frac{A_i}{A} = \Gamma_i \cdot \frac{1}{1-\epsilon} , \qquad (2)$$

with $N_i^{chains}$ the number of polymer chains initially adsorbed. As shown by Millet [13-15] and our group [17], PAAH-α–$C_n$ layers form brush-like layers, where hydrophobic anchors adsorb at the interface, and hydrophilic loops stretch into the solution.

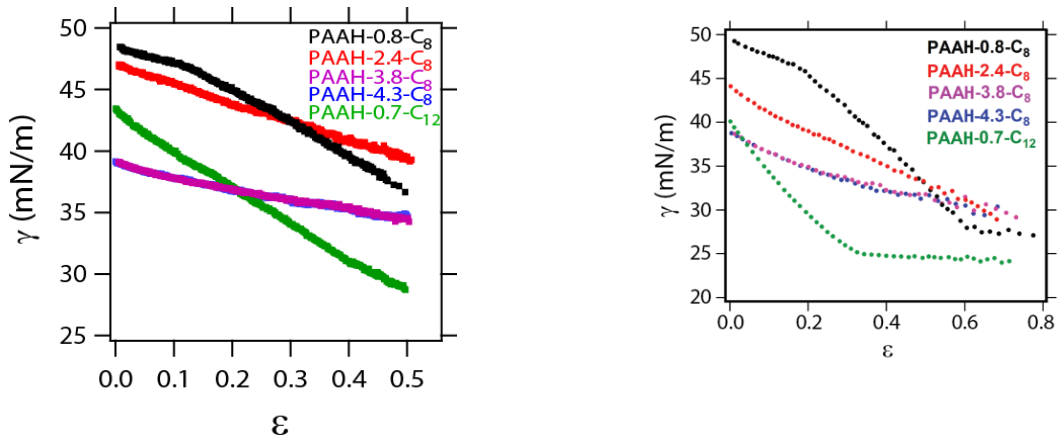

(a) (b)

*Figure 1: Evolution of the surface tension of PAAH-α-Cn layers during compression at a rate of (a) 0.007 s$^{-1}$ and (b) and 0.1 s$^{-1}$.*

According to Aguié-Béghin *et al*. [2] the surface pressure $\Pi$ of a copolymer layer in a regime where the hydrophobic parts are adsorbed to the air/water interface and the hydrophilic parts form stretched loops in solution reads

$$\Pi = \gamma_0 - \gamma = \frac{k_B T N \Gamma}{N_L}, \quad (3)$$

where $\gamma_0$ and $\gamma$ are the interfacial tensions of pure water and of the solution, $N$ is the number of monomers per polymer chain and $N_L$ is the number of monomers per loop. In our case we assume that a polymer chain composed of $N$ monomers typically contains $\alpha N$ grafts, and $\alpha N$ loops, which contain $1/\alpha$ monomers each. Therefore eq. (3) becomes

$$\Pi = \gamma_0 - \gamma = k_B T \, \alpha N \, \Gamma. \quad (4)$$

If the number of adsorbed entities is constant during the deformation, we can predict the evolution of the surface tension, $\gamma_{\text{no desorption}}$, according to the following equation

$$\gamma_{no\ desorption} = \gamma_0 - k_B T \frac{\alpha N \Gamma_i}{1-\epsilon} = \gamma_0 - \frac{\gamma_0 - \gamma_i}{1-\epsilon}, \quad (5)$$

with $\gamma_i$ the initial surface tension before compression. $\gamma_{no\ desorption}$ can also be written

$$\gamma_{no\ desorption} = \gamma_0 - \Pi_{no\ desorption}, \quad (6)$$

with
$$\Pi_{no\ desorption} = \frac{\gamma_0 - \gamma_i}{1-\epsilon}, \quad (7)$$

which is the surface pressure obtained if no desorption of the chains occurs. In Figure 2 we represent the measured surface tension, $\gamma$, as a function of $\Pi_{no\ desorption}$. The red dashed lines are the predictions from Eq. 6. A compression with desorption would lead to higher surface tensions, *i.e.* to variation curves located above that prediction.

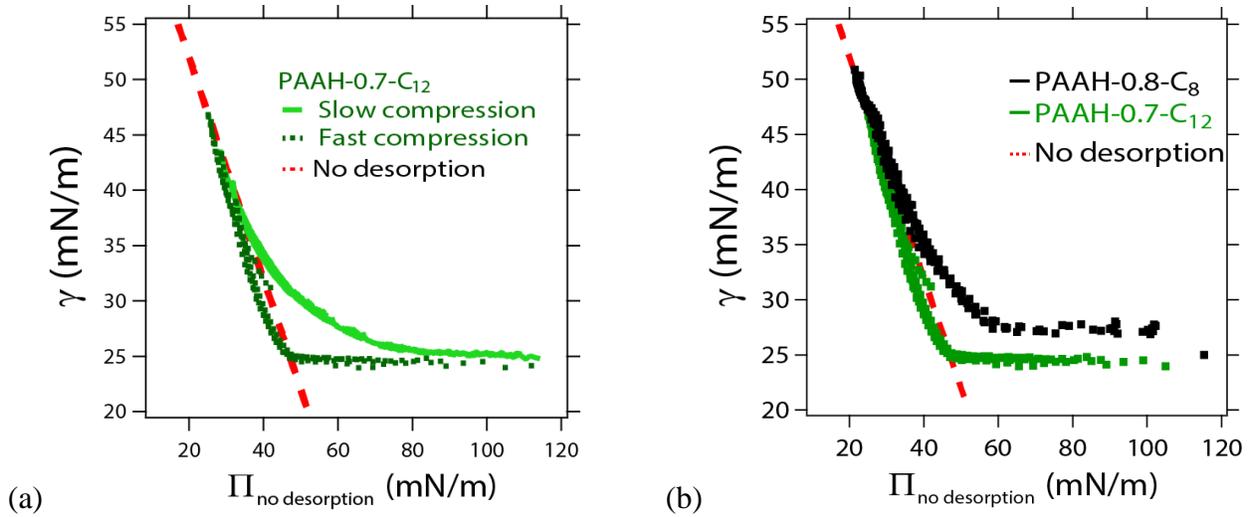

*Figure 2: Surface tension as a function of the reference surface pressure (see eq. 7) if the number of adsorbed molecules stays constant during the compression of the interface. (a) Compression of PAAH-0.7-$C_{12}$ at compression rates of 0.007 $s^{-1}$ and 0.1 $s^{-1}$. (b) Compression of PAAH-0.8-$C_8$ and PAAH-0.7-$C_{12}$ at a rate of 0.1 $s^{-1}$.*

The compression curves shown in Figures 2a and 2b present two regimes. At low compression and thus low $\Pi_{no\ desorption}$, the surface tension superimposes well with the prediction of eq. 6, which indicates that there is no desorption in this regime. At larger compressions and thus high $\Pi_{no\ desorption}$, the surface tension saturates to a plateau indicating that some desorption or partial desorption of molecules occurs, and that the surface excess reaches a maximal value due to possible steric/free-enthalpic constraints at the surface. We suggest that the desorption process concerns the hydrophobic anchors, as previous investigations [13-15, 17] showed that the hydrophilic AAH blocks form loops extending in solution while the $C_n$ grafts adsorb at the interface. For a faster compression of the PAAH-0.7-$C_{12}$ layers the desorption occurs at a larger deformation and the transition between the two regimes is sharper (Figure 2a). Moreover, when comparing $C_8$ and $C_{12}$ anchors, we find that the surface tension remains higher for $C_8$ grafts than for $C_{12}$ grafts during compression in agreement with the lower anchoring ability of the smaller grafts. The driving mechanism for the desorption is the increase of chemical potential of the chains in the brush. In References 10 and 30, it was shown that the rate of desorption decreases with the length of the alkyl tail hence for a given compression speed, we therefore expect that $C_8$ grafts will desorb faster than $C_{12}$ grafts, consistently with the observation that for $C_8$ grafts the surface tension leaves the "no-desorption" curve at a lower deformation than for $C_{12}$ grafts.

Moreover for a given alkyl anchor length, if the compression is faster than the rate of desorption, we expect to follow the "no-desorption" curve while an infinitely low compression speed the desorption of the grafts should favor a constant interfacial tension, consistently with our results which show that the surface tension remains closer to the "no-desorption" curve for a higher compression speed.

In Figure 3a we study the influence of the grafting degree on the compression behavior for the PAAH-α-$C_8$ chains. We observe that the more grafted the chain is, the lower the initial surface tension is, as a result of a better affinity of the chains with the interface. Remarkably the surface tension at large compression is further above the "no-desorption" curve for the chains with a higher grafting degree than for the less grafted chains.

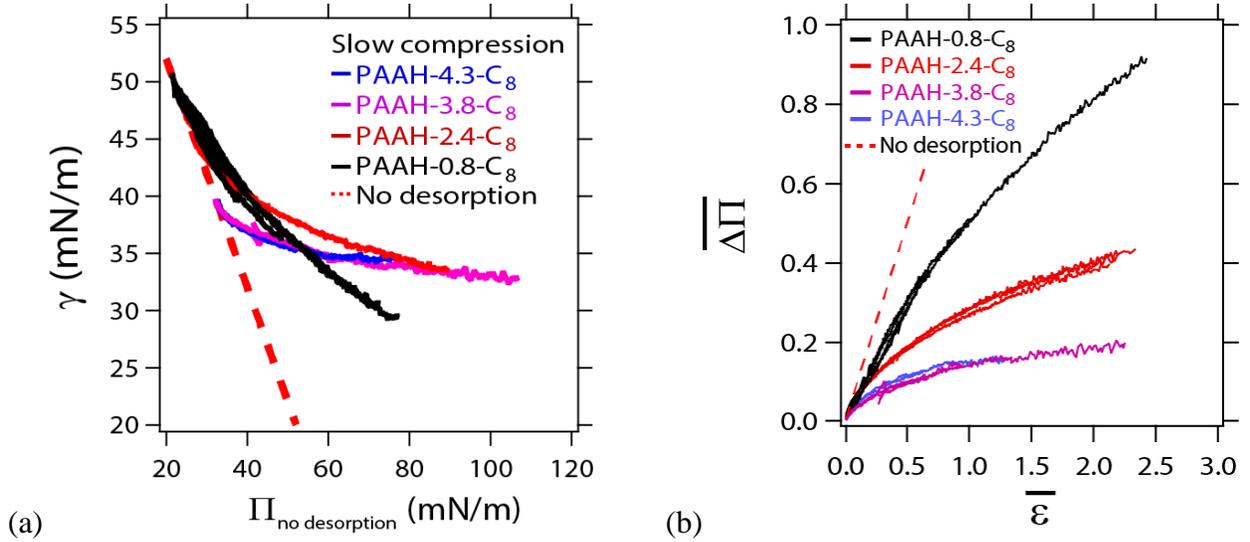

(a)  (b)

*Figure 3: (a) Surface tension as a function of the surface pressure predicted for a constant number of adsorbed monomers (see eq. 7) during the compression of the interface covered by monolayers of different grafting densities.; (b) Relative change in surface pressure as a function of the compression parameter $\bar{\epsilon}$ (see eq. 9) for the different grafted monolayers.*

To rationalize this counter-intuitive finding, we provide in Figure 3b another way of plotting the results. We define the following dimensionless numbers: the relative change in surface pressure $\overline{\Delta \Pi}$, and the compression parameter $\bar{\epsilon}$ defined as follows:

$$\overline{\Delta \Pi} = \frac{\Pi - \Pi_i}{\Pi_i}, \tag{8}$$

and

$$\bar{\epsilon} = \frac{A_i - A}{A} = \frac{1}{1-\epsilon} - 1, \tag{9}$$

The compression parameter $\bar{\epsilon}$ increases during compression starting from zero.

Using Equations (7, 8, 9), the relative change in the surface pressure in the case of no desorption can be written as:

$$\overline{\Delta\Pi_{\text{no desorption}}} = \frac{\Pi_i \cdot \frac{1}{1-\epsilon} - \Pi_i}{\Pi_i} = \bar{\epsilon}. \tag{10}$$

In Figure 3b, we plot the relative change in surface pressure versus the compression parameter $\bar{\epsilon}$ for slow compression of polymer PAAH-α-C$_8$, with different grafting densities. The red dashed line represents the theoretical prediction of Equation (10) corresponding to the case of no desorption. We systematically find that $\overline{\Delta\Pi} < \bar{\epsilon}$. The more grafted the chains, the stronger the deviation from Equation (10), which indicates a stronger desorption.

Taking into account the fact that in these pseudo-brushes the hydrophobic grafts adsorb in the interface while the hydrophilic blocks form loops away from the interface, one can reasonably argue that the number of hydrophobic anchors directly determines the length of the loops. Assuming in first approximation that all grafts adsorb at the interface, a loop between two adsorbed grafts has a length 1/α in number of monomers, then the ideal entropic spring constant which resists the compression scales as α per loop and thus N.α² per chain. Thus when compressing the layer, higher alpha leads to higher elastic penalty and thus faster desorption/diffusion rate.

*Large dilatations of the layers*

We now probe the effect of large dilatations. We plot in Figure 4 the measured surface tension as a function of $\Pi_{\text{no desorption}}$ (see eq 7) during the dilatation. The surface tensions obtained for the chains with the lowest degrees of grafting ($\alpha = \frac{0.7}{100}$) follow the prediction of Equation (6) which corresponds to a constant number of adsorbed species, ie the absence of any adsorption or desorption of the chains. This indicates that as the interface is dilated (from right to left on the x-axis, due to the minus sign in ε), the surface coverage decreases but the total number of adsorbed entities remains constant. The dilatation occurs too quickly to allow for significant adsorption during the deformation. These observations are consistent with the fact that for the chains with low degrees of grafting, all the hydrophobic grafts of the adsorbed chains are already adsorbed before dilatation, and that diffusion of new chains toward the interface is to slow to decrease the surface tension. Note that the AAH monomers comprised in the loops of the brush may in principle also adsorb upon dilatation, but we know from previous studies that the adsorption of AAH monomers does not lead to significant decrease of the surface tension [17, 31]. For chains with larger degrees of grafting, $\alpha = 2.4/100 - 3.8/100$, the interfacial tension is slightly lower than the dashed "no-adsorption" line. This indicates that new molecules or entities have time to adsorb while the interface is expanded (from right to left on x-axis). We suggest that some unadsorbed grafts may remain in the vicinity of the interface hence can adsorb quickly during interfacial dilatation.

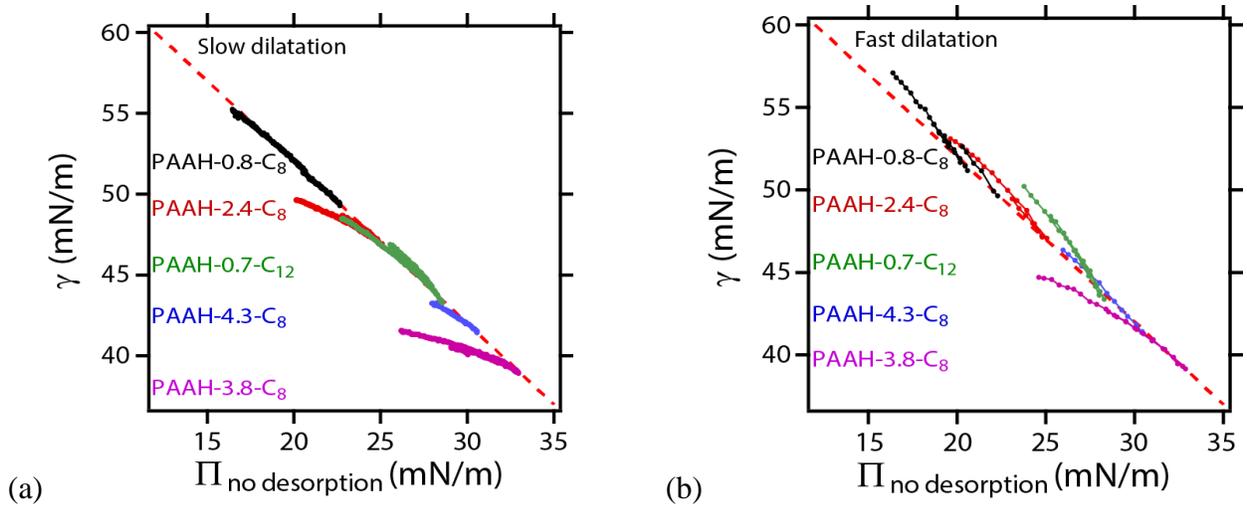

*Figure 4: Surface tension as a function of reference surface pressure (see Eq. 7) during the dilatation of the interface covered by monolayers of chains with different grafting densities and graft lengths, as indicated. (a) Dilatation rate is 0.007 Hz; (b) dilatation rate is 0.1 Hz.*

**Conclusion**

In this article, we investigate the behaviour of PAAH-$\alpha$-$C_n$ polymer layers at air-water interfaces when submitted to large interfacial compressions and dilatations. By comparing the interfacial tension with a prediction obtained for a constant number of adsorbed entities, the compression experiments enable to probe the forced desorption dynamics of these layers. We show that longer alkyl grafts desorb more slowly resulting in lower transient interfacial tensions than for shorter grafts. Moreover faster compressions lead to lower transient interfacial tensions because the grafts have less time to desorb. Furthermore, we observe more desorption for higher grafting densities, which is a counter intuitive result. We attribute this result to the fact that for high grafting densities, the number of monomers per loop is lower, hence the elastic penalty caused by the compression is higher. The dilatation experiments show that the number of adsorbed anchors remains constant in the case of chains with low grafting densities suggesting that all the grafts of the adsorbed chains are initially adsorbed. However, chains with the highest degrees of grafting seem to show some adsorption during the dilatation. Hence, for these chains there may be unadsorbed grafts remaining in the vicinity of the interface, which may adsorb quickly to the interface upon dilatation.

**Acknowledgements**

The authors acknowledge ANR JCJC INTERPOL for financing this work.